\documentclass[journal]{IEEEtran}

\usepackage{graphicx}
\usepackage{color}
\usepackage{bbold}
\usepackage{amsmath,amssymb}
\usepackage{multirow}
\usepackage{bm}
\usepackage{cite}
\usepackage{flushend}

\usepackage{caption}
\usepackage{subcaption}

\usepackage[acronym]{glossaries}
\newacronym{gam}{GAM}{Generalised Additive Model}
\newacronym{gamlss}{GAMLSS}{Generalised Additive Model for Location, Scale and Shape}
\newacronym{nwp}{NWP}{Numerical Weather Prediction}
\newacronym{gpd}{GPD}{Generalised Pareto Distribution}
\newacronym{gspg}{GSP Group}{Grid Supply Point Group}
\newacronym{gb}{GB}{Great Britain}
\newacronym{tso}{TSO}{Transmission System Operator}

\newcommand{\blue}[1]{#1}

\begin{document}
\title{Probabilistic Forecasting of Regional Net-load with Conditional Extremes and Gridded NWP}
%
%
%

\author{Jethro~Browell~
and~Matteo~Fasiolo
\thanks{J. Browell is with the Department of Electronic and Electrical Engineering, University of Strathclyde, Glasgow, UK, (www.jethrobrowell.com).}
\thanks{M. Fasiolo is with the School of Mathematics, University of Bristol, UK.}
}

%



\maketitle

\begin{abstract}
The increasing penetration of embedded renewables makes forecasting net-load, consumption less embedded generation, a significant and growing challenge. Here a framework for producing probabilistic forecasts of net-load is proposed with particular attention given to the tails of predictive distributions, which are required for managing risk associated with low-probability events.
Only small volumes of data are available in the tails, by definition, so estimation of predictive models and forecast evaluation requires special attention.
%
We propose a solution based on a best-in-class load forecasting methodology adapted for net-load, and model the tails of predictive distributions with the Generalised Pareto Distribution, allowing its parameters to vary smoothly as functions of covariates. The resulting forecasts are shown to be calibrated and sharper than those produced with unconditional tail distributions. In a use-case inspired evaluation exercise based on reserve setting, the conditional tails are shown to reduce the overall volume of reserve required to manage a given risk. Furthermore, they identify periods of high risk not captured by other methods. The proposed method therefore enables user to both reduce costs and avoid excess risk.

\end{abstract}


\IEEEpeerreviewmaketitle

\section{Introduction}

\IEEEPARstart{T}{he} nature of distribution networks is changing fundamentally as the penetration of generation, storage and flexible loads connected to them has grown over recent decades. Transmission networks are no longer simply supplying energy to meet the demand of connected consumers, but the net-load of consumption less production. Some distribution networks may even export energy to the transmission system during times of high renewable output. This complexity makes predicting future net-load far more challenging than conventional load forecasting, while such forecasts are increasingly valuable to both system operators and energy market participants. Furthermore, probabilistic forecasts have an important role to play in communicating forecast uncertainty which is impacted by the composition of net-load at any given time.

Our focus here is twofold: first, how to incorporate embedded generation in probabilistic load forecasts through use of weather forecasts and limited knowledge of installed capacities; and second, how to best describe the tails of predictive distributions which, while being valuable for many users, is challenging and has received little attention in the literature. 
While domestic solar is seen as the  main challenge in some areas, in Great Britain and elsewhere small-scale wind, MW-scale solar arrays, batteries, and small thermal plant, such as waste-to-power plants and CHP, are often `embedded' and only visible to network operators as negative load. 
\blue{In 2018, the last year of the case study presented in this paper, net-load on the \gls{gb} electricity transmission system (also called Transmission System Demand) ranged from 18GW to 50GW, with 13GW and 6GW of embedded solar and wind capacity, respectively, and a further 12GW of embedded conventional plant, mostly gas-fuelled. Periods of low load and high wind and solar generation present a significant challenge to the \gls{tso} due to their lack of visibility and control-ability of embedded generation~\cite{Drew2019}.}

There is a large literature on probabilistic load forecasting and such forecasts are increasingly being adopted by users in industry. Hong \textit{et al} review the topic in~\cite{Hong2016a} and highlight net-load forecasting as a new problem. Since then, day-ahead net-load forecasting at neighbourhood level has been studied in~\cite{Kobylinski2020}, but only deterministic predictions are considered. Probabilistic forecasts in the form of intervals are studied in~\cite{Meer2018} and \cite{Pierro2020} in the context of increasing embedded solar. While the former focuses on customer aggregation, the latter considers reserve scheduling and reaches the conclusion that greater reserves are required to manage increased uncertainty.

Regarding specific forecasting methodology, Hong \textit{et al} conclude that that ``a universally best technique simply does not exist''.
Indeed, the implementation of prediction algorithms is of comparable importance to the choice of algorithm itself, with performance sensitive to hyper-parameter tuning, data preparation, and feature engineering, for example.
That said, a high-level observation can be made that non-parametric probabilistic forecasting, and quantile regression in particular, is generally favoured~\cite{Hong_2016,Li2020}. When considering net-load, it is reasonable to expect that innovations in wind and solar generation forecasting may enhance the accuracy of net-load forecasts.
For instance, sky cameras has been used for 10--30 minute-ahead net-load forecasting with high penetrations of embedded solar~\cite{Chu2017}.
Additionally, leveraging a grid of \glspl{nwp}, rather than a single point, can improve day-ahead wind and solar forecasts markedly~\cite{Andrade2017}, a concept we adapt for net-load forecasting here, following the holistic variable selection approach of Xie and Hong~\cite{Xie2018}.

For many users, the accuracy of extreme probabilities is an important aspect of probabilistic forecasting, i.e. the tails of predictive distributions. Quantile regression is widely used for probabilistic forecasting as it makes no assumptions about the shape of predictive distributions, but is not well suited to estimating conditional extremes as data is increasingly sparse in the tails, so special treatment is required. While there are well established practices modelling extreme events \blue{for planning purposes \cite{Moreno2020}, such heat waves, the extremes of short-term forecast errors and uncertainty have received little attention.
Panteli \textit{et al} discuss the importance of short-term forecasts of extremes in power system resilience in \cite{Panteli2017} and highlighted this as an area where development of advanced tools is required. It should be noted that severe weather events are typically well forecast in the short-term, and that extremes of short-term forecast errors/uncertainty do not necessarily correspond to extreme weather.}

Extreme value theory provides a theoretical framework for this task, particularly the peaks-over-threshold method and the resulting \gls{gpd}~\cite{Coles2001}. The \gls{gpd} has recently been applied to the tails of probabilistic wind power forecasts and shown to be effective where quantile regression fails~\cite{Goncalves2021}. We propose applying a similar technique for net-load here, but, while in~\cite{Goncalves2021} the \gls{gpd} parameters are fixed, we allow them to depend on covariates along the lines of~\cite{Youngman2020}.
In the case of net-load, we suspect that, similarly to the central portion of the predictive distribution, the tails will depend on behavioural and weather effects. \blue{In the remainder of this paper we use `extreme' refer to probability levels where standard quantile regression is not suitable.}
%

Energy forecasts of varying levels of sophistication play an important role in reserve setting and procurement, a process that impacts the cost of operating a power system and its reliability~\cite{Matos2011,Holttinen2012,Rajbhandari2016}. Requirements for reserve services are set to meet a given risk appetite based on the probability of events such as unplanned plant outages and large forecast errors.
%
In a risk-neutral setting, the relevant probability is the ratio of the marginal reserve cost to value of lost load, typically 0.01\%--0.25\%, which is approximately the same range bounded by common heuristics 1 hour and 1 day-per-year. Forecasts of net-load are therefore required at these probability levels.
The impact of solar penetration on net-load and reserve setting was recently explored in \cite{Pierro2020} concluding that the need for reserve increases with solar penetration based on 95\% prediction interval width. Here, we develop a similar approach to quantify the value of forecast improvement in terms of volume of reserve required to manage forecast uncertainty and compare regions with differing types and capacities of embedded generation.

This paper contributes a novel model structure for probabilistic net-load forecasts in the presence of embedded wind and solar generation, including a methodology for forecasting extreme quantiles; an evaluation framework for the tails of predictive distributions previously lacking; and the augmentation of net-load forecasts using features derived from gridded \gls{nwp} data. These methods are demonstrated on large dataset which is published along with the paper, and their performance is evaluated in terms of both statistical metrics and practical value for procuring reserve services. 

\section{Net-load Forecasting}
\label{sec:net-load}

Electricity load forecasting has been the focus of much research resulting in a range of models which may be considered `best-in-class' for this task~\cite{Hong2016a}. For the purposes of exploring the value of gridded \gls{nwp} and estimation of extreme quantiles, we have selected one such approach for producing probabilistic forecasts of non-extreme quantiles, that of~\cite{Gaillard2016} based on \glspl{gam} and linear quantile regression, which won the load track of GEFcom2014. We also consider \textit{Tao's Vanilla Benchmark} as an additional reference model~\cite{Hong2012}, extended using linear quantile regression in the same way as the aforementioned \glspl{gam}.

\subsection{Generalised Additive Models for Location Shape and Scale}
\label{sec:gamlss}


\begin{table}
\caption{Description of forecasting models and feature sets used by each. Feature sets are detailed in Table~\ref{tab:features}.}
    \centering
    \begin{tabular}{l|p{6cm}} \hline
         Model & Description \\ \hline
         Vanilla-T & Tao's Vanilla Benchmark \cite{Hong2012} \\
         GAM-T & GAM with date/time and temperature features only \\
         Vanilla-Point & Tao's Vanilla Benchmark with addition of linear wind speed and solar irradiance features \\
         GAM-Point  & GAM-T with additional standard \gls{nwp} features \\
         GAM-Grid  & GAM-Point with additional features derived from gridded \gls{nwp} \\ \hline
    \end{tabular}
    \label{tab:models}
\end{table}

Let $y_t$ be the net-load at time $t$ and indicate with $F_t(y_t)$ its cumulative probability distribution. In a distributional regression context, $F_t(y_t)$ is modelled via a parametric model, $F(y_t|\bm \theta_t)$, where $\bm \theta_t$ is an $m$-dimensional vector of parameters. In \gls{gamlss} \cite{Rigby2005} the elements $j=1,...,m$ of $\bm \theta_t$ are modelled via
\begin{equation} \label{eq:basicGAM}
    g_j(\theta_{j,t})=\mathbf{A}_{j,t} \bm{\beta}_j + \sum_{i} f_{j,i}({\bm x}^{S_{j,i}}_t), \;\;\; \text{for} \;\;\; j = 1, \dots, m,
\end{equation}
where $g_j$ is a monotonic function, $\mathbf{A}_{j,t}$ is the $t$-th row of the design matrix $\mathbf{A}_j$, $\bm \beta_j$ is a vector of regression coefficients, $\bm x_t$ is a $d$-dimensional vector of covariates and $S_{j,i} \subset \{1, \dots, d\}$. Hence, if $S_{j,i} = \{1, 3\}$, then following our notation ${\bm x}_{t}^{S_{j,i}}$ is a two dimensional vector formed by the first and third element of $\bm x_t$. Each $f_{j,i}$ is a smooth function, built via
\begin{equation}
    f_{j,i}(\bm x^{S_{j,i}}) = \sum_{k=1}^{K_{j,i}} b^{ji}_k (\bm x^{S_{j,i}}) \beta_k^{ji},
\end{equation}
where $b^{ji}_k$ are spline basis functions of dimension $\vert S_{j,i} \vert$, while $\beta_k^{ji}$ are regression coefficients. The smoothness of each $f_{j,i}$ is controlled via ridge penalties, the definition of smoothness being dependent on the type of effect and penalty being used. In this work we will use the term generalised additive model (GAM) to indicate a restricted \gls{gamlss} model where only one element of $\bm \theta_t$ is modelled as in (\ref{eq:basicGAM}), while the other elements are kept constant. See \cite{Wood2017} for a detailed introduction to GAM/GAMLSS models, smoothing splines bases and penalties.

\subsection{Multiple Quantile Regression}
\label{sec:quantile_regression}

Quantile regression enables the construction of a non-parametric predictive distribution by from multiple conditional quantiles. While \cite{Fasiolo2020} provides method for fitting quantile GAMs, these are too computationally demanding for the forecasting application considered here, which requires many complex model fits. 
Therefore, we apply linear quantile regression to the residuals of the main \gls{gam} (or Vanilla) model with a reduced feature set, $\mathbf{B}_t$. The predictive quantile for probability level $\alpha$ is given by
\begin{eqnarray}
    q_{\alpha,t} &=& \hat{y}_t + \mathbf{B}_t \beta_\alpha \quad ,
    \label{eqn:qreg}
\end{eqnarray}
where $\hat{y}_t$ is the deterministic forecast, which is an estimate of the conditional expectation 
produced by a Gaussian \gls{gam} (or Vanilla model) and $\beta_\alpha$ are model parameters to be estimated.
The features $\mathbf{B}_t$ may be either exogenous variables or, following~\cite{Gaillard2016}, the design matrix of linear terms from the fitted \gls{gam} model.

The predictive distribution $\hat{F}_{\text{QR},t}(y_t)$ is constructed by estimating multiple quantiles at a range of probability levels $\alpha_L,...,\alpha_R$ and interpolating between them to produce a continuous cumulative distribution function. An obvious drawback of this approach is that the predictive distribution is not defined for probabilities below $\alpha_L$ or above $\alpha_R$. For some applications this may be of no consequence, but where full support is required, such as sampling, modelling dependency using copulas, or setting reserves based on extreme quantiles, a solution is needed. One simple solution is to confine distribution within a upper and lower boundaries by assigning values to $q_{0,t}$ and $q_{1,t}$ and including them in the interpolation scheme. If the target variable is bounded by physical or other known restrictions, then this may be justified provided that $\alpha_L$ is close enough to 0 (and $\alpha_R$ to 1) for the resulting interpolation to result in a distribution of reasonable quality. The alternative is to specify a parametric function describing the tails of the predictive distribution, and this is the subject of Section~\ref{sec:conditional_extremes}.

\subsection{Weather and Date/time Features}
\label{sec:weather_features}

Conventional load forecasting typically considers the impact of temperature on electricity consumption, which is largely heating and cooling demand, and also influenced to a lesser extent by wind chill, precipitation, and illumination. Furthermore, a single-valued weather feature is typically assumed to be representative for the geographic region of interest. This may be derived from one or more weather forecast/measurement locations, typically meteorological stations close to population centres, and/or multiple variables, as in the case of so-called \textit{composite weather variables}. 

In order to explore the value of weather features targeting embedded wind and solar generation, and more sophisticated treatment of weather data in general, we compare \gls{gam} models with the same basic structure but with different sets of weather features. \glspl{nwp} are produced on a regular spatial grid and therefore contain information beyond `average' conditions for given region. Statistics derived from gridded \gls{nwp} such as spatial standard deviation, min, max and so on, encode potentially valuable information for predicting net-load in a region, as demonstrated for wind and solar production in~\cite{Andrade2017a}, so we consider including them here in net-load forecasting models.

We consider a progression of models with increasingly sophisticated weather features \blue{combining best-practice from load and generation forecasting}: We first consider models that only include calendar and point temperature features (labeled `-T'). Point temperatures forecasts are those for the single \gls{nwp} cell corresponding to highest population density in a region. Next, we add the spatially-averaged wind speed and solar irradiance forecasts for each region combined with embedded wind and solar capacity, which may change significantly over time (labeled `-Point'). Finally, we include statistics such as spatial standard deviation, min and max derived from gridded \gls{nwp} (labeled `-Grid'). Full lists of models and features are provided in Tables \ref{tab:models} and \ref{tab:features}, respectively.

\blue{Date/time features play a central role in load forecasting to capture diurnal, weekly and annual seasonality. Here we use the GAM framework to estimate a smooth time-of-day profile for nine day-types: each day of the week, public holidays and school holidays following~\cite{Ziel2018}. We also include interactions between time-of-day and weather effects as these are known to elicit a different response depending on the time of day~\cite{Hong2012}. Autoregressive terms are also a common feature of short-term load forecasting and can capture diurnal and weekly seasonality as well as level-changes in load. However, we found them to be detrimental here, possibly due to the impact of embedded generation polluting the seasonal cycles, and instead opted to include a two-week rolling mean of net-load to capture level-changes with less sensitivity to individual lagged observations.}

\begin{table}
\setlength{\tabcolsep}{2pt}
\caption{\blue{List of explanatory variables, type (L=linear, D=dummy, P($n$)=polynomial of order $n$, S=smooth, BS=bivariate smooth), and details of the cubic spline basis dimension $K$ used in \gls{gam} modelling. Features labeled $\checkmark_q$ are also as linear features in quantile regression, $\mathbf{B}_t$ in Equation \eqref{eqn:qreg}. Calculation of weather features is described in Section~\ref{sec:weather_features}.}}
    \centering
    \begin{tabular}{p{4.3cm}|c|c|c|c|c|c} \hline
         Terms & Type & \rotatebox[origin=c]{90}{Vanilla-T}
         & \rotatebox[origin=c]{90}{GAM-T}& \rotatebox[origin=c]{90}{Vanilla-Point}& \rotatebox[origin=c]{90}{GAM-Point}& \rotatebox[origin=c]{90}{GAM-Grid}\\ \hline
         Linear trend (time since 2014-01-01) & L & $\checkmark$ & & $\checkmark$ & & \\
      Clock-time (local time zone) & D & $\checkmark_q$ & & $\checkmark_q$ & & \\
      Day-type \{Mon,Tue,...,Sun, Public Holiday\} & D & $\checkmark_q$ & $\checkmark_q$ & $\checkmark_q$ & $\checkmark_q$ & $\checkmark_q$ \\
         Trend & P($2$) &  & $\checkmark$ & & $\checkmark$ & $\checkmark$ \\
         Two-week rolling average net-load & L &  & $\checkmark$ & & $\checkmark$ & $\checkmark$ \\
         Second-order Fourier (annual) & L &  & $\checkmark$ & & $\checkmark$ & $\checkmark$ \\
         School Holiday (region-specific) & D &  & $\checkmark$ & & $\checkmark$ & $\checkmark$ \\
         Clock-time (decimal hour in local time zone) & S, $K$=35 &  & $\checkmark$ & & $\checkmark$ & $\checkmark$ \\
         Clock-time by day-type & S, $K$=30 &  & $\checkmark$ & & $\checkmark$ & $\checkmark$ \\
         Clock-time by school holiday & S, $K$=20 &  & $\checkmark$ & & $\checkmark$ & $\checkmark$ \\
         Point temperature & P($3$) &  $\checkmark_q$ & & $\checkmark_q$ & & \\
         Point temperature by clock-time & P($3$) & $\checkmark$ & & $\checkmark$ & & \\
         Point temperature by month & P($3$) &  $\checkmark$ & & $\checkmark$ & & \\
         Point temperature & S, $K$=35 &    & $\checkmark$ & & $\checkmark$ & $\checkmark$ \\
         48-hour rolling mean point temperature & S, $K$=35 &   & $\checkmark$ & & $\checkmark$ & $\checkmark$ \\
         Population-weighted temperature & S, $K$=35 &   & & & & $\checkmark_q$ \\
         48-hour rolling mean population-weighted temperature & S, $K$=35 &   & & & & $\checkmark$ \\
         Population-weighted temperature and clock-time & BS, $K$=17 & & & & & $\checkmark$ \\
         Mean irradiance scaled by solar capacity & L &   &  & $\checkmark_q$ &  & \\
         Mean irradiance scaled by solar capacity & S, $K$=5 &   & & & $\checkmark$ & $\checkmark$ \\
         Mean irradiance and 10m wind speed & BS, $K$=17 & & & & $\checkmark$ & $\checkmark$ \\
         Max irradiance scaled by solar capacity & S, $K$=5 & & & &  & $\checkmark$ \\
         Standard deviation of irradiance scaled by solar capacity & S, $K$=5 & & & &  & $\checkmark_q$ \\
         Max cloud cover scaled by solar capacity & S, $K$=5 & & & &  & $\checkmark$ \\
        
    Mean 10m wind speed & L & & & & $\checkmark$ & $\checkmark$ \\
    Mean 100m wind speed & L & & & $\checkmark_q$ & & \\
    Mean 100m wind speed by embedded wind capacity & S, $K$=20 & & & & $\checkmark_q$ & $\checkmark_q$ \\
    Standard deviation of 100m wind speed & L & & & & & $\checkmark_q$ \\
    Day-ahead electricity price (N2EX) and clock-time & BS, $K$=17 & & & & $\checkmark$ & $\checkmark$ \\
    Mean precipitation & S, $K$=5 & & & & $\checkmark$ & $\checkmark$ \\
    Mean precipitation and clock-time & BS, $K$=17 & & & & $\checkmark$ & $\checkmark$ \\
    Standard deviation of precipitation & S, $K$=5 & & & & & $\checkmark$ \\ \hline
    \end{tabular}
    \label{tab:features}
\end{table}

\section{Conditional Extremes}
\label{sec:conditional_extremes}

Quantile regression is not well suited to estimating extreme conditional quantiles due to the sparsity of data by definition, which is further exacerbated by conditioning on exogenous variables~\cite{Koenker2008}.
The peaks-over-threshold method provides a parametric estimator for extreme quantiles, namely the scaled \gls{gpd}, given by
\begin{equation}
    F_\text{GPD}(y|\sigma,\xi) = \begin{cases}
1 - \left(1+ \frac{\xi y}{\sigma}\right)^{-1/\xi} & \text{for }\xi \neq 0, \\
1 - \exp \left(-\frac{y}{\sigma}\right) & \text{for }\xi = 0,
\end{cases}
\end{equation}
where the support of $Y$ is $y \geqslant 0 $ when $\xi \geqslant 0 \,$, and $ 0 \leqslant y \leqslant - \sigma /\xi $  when $ \xi < 0$~\cite{Coles2001}.
The parameters of the \gls{gpd} can be allowed to vary with the covariates as described in Section~\ref{sec:gamlss}, thus leading to the following GAMLSS model
\begin{eqnarray}
    g_1(\sigma_t) &=& \mathbf{C}_{1,t} \bm{\beta}_1 + \sum_i f_{1,i}(x_{i,t}) + ... \\
    g_2(\xi_t) &=& \mathbf{C}_{2,t} \bm{\beta}_2 + \sum_i f_{2,i}(x_{i,t}) + ...
\end{eqnarray}
which may be estimated using the general GAM methods of~\cite{wood2016smoothing}.

\begin{figure}
    \centering
    \includegraphics[width=\columnwidth]{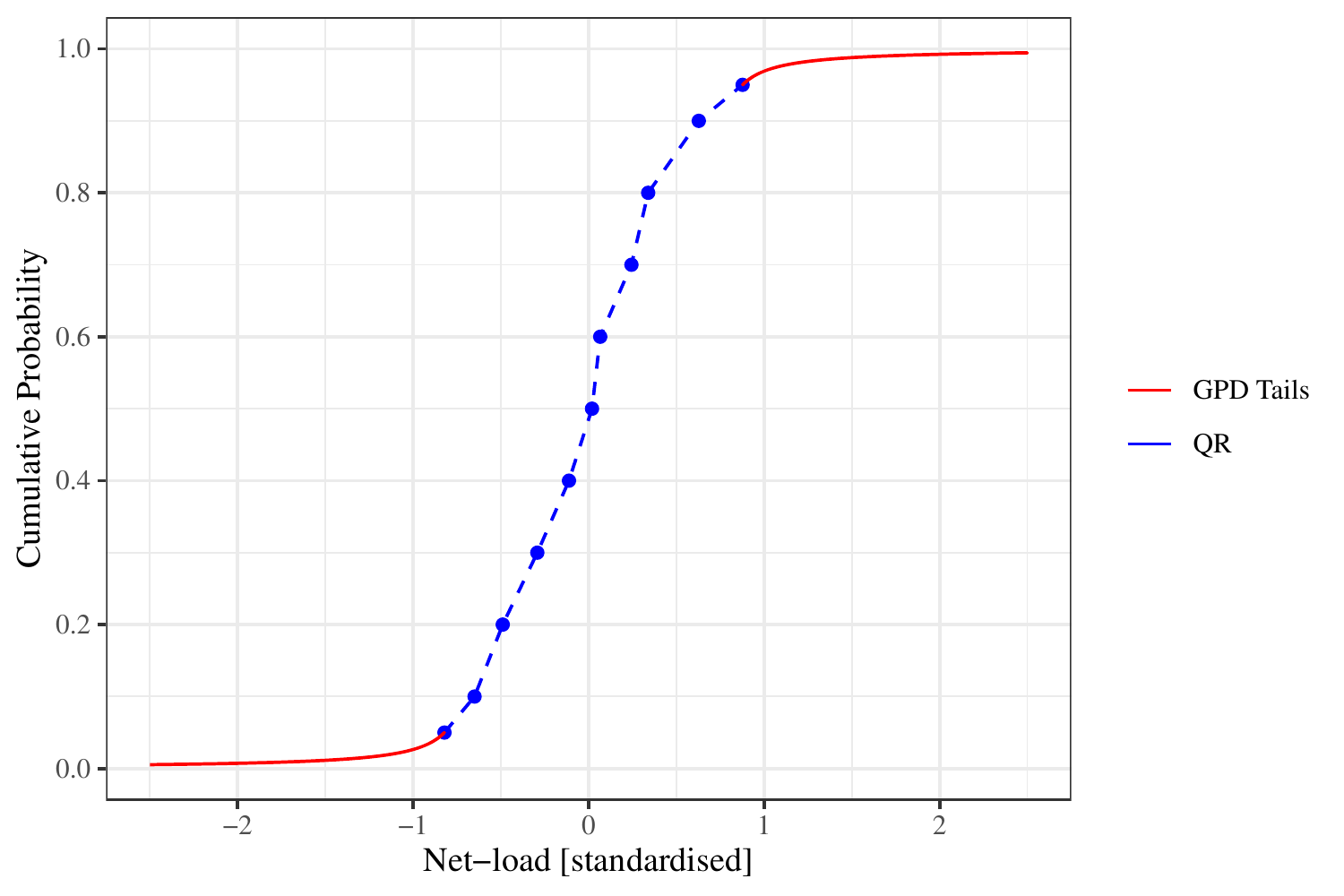}
    \caption{Illustration of predictive distribution comprising quantile regression (QR) between $\alpha_L=0.05$ and $\alpha_R=0.95$ and \gls{gpd} tails described in Equation~\eqref{eqn:qr+gpd}.}
    \label{fig:gpd_illustration}
\end{figure}

Integrating the \gls{gpd} model with quantile regression provides full support for $\hat{F}$, which now comprises conditional quantiles between probability levels $\alpha_L$ and $\alpha_R$, and two \glspl{gpd} for probabilities $\alpha<\alpha_L$ and $\alpha>\alpha_R$. The predictive distribution  $\hat{F}(y|\bm{x})$ may be written
\begin{equation}
    \begin{cases}
    \alpha_L F_\text{GPD}(q_{\alpha_L}-y|\sigma_L({\bm x}),\xi_L({\bm x}))
    \hfill~\text{for }y < q_{\alpha_L}, \\
    \hat{F}_\text{QR}(y|{\bm x})
     \hfill~\text{for } q_{\alpha_L} \le y \le q_{\alpha_R}, \\
    \alpha_R + (1-\alpha_R) F_\text{GPD}(q_{\alpha_R}+y|\sigma_R({\bm x}),\xi_R({\bm x}))
    \hfill~\text{for } y > q_{\alpha_R},
\end{cases}
\label{eqn:qr+gpd}
\end{equation}
and illustrated in Figure~\ref{fig:gpd_illustration}.

The choice of $\alpha_L$ and $\alpha_R$, which specify the points at which quantile regression meets the parametric tails, requires some consideration and depends on both data volume and complexity of the models involved (e.g. number of covariates). We suggest that quantile regression should be extended as far into the tail as possible, i.e. until $q_{\alpha_L}$ and $q_{\alpha_R}$ become unreliable or it becomes difficult to verify their reliability, based on a cross-validation exercise. This is the procedure we employ in the case study presented in Section \ref{sec:case_study}, but alternative strategies may be imagined.

In the context of net-load forecasting, we hypothesise that the tails of the predictive distribution depend on multiple factors, including time-of-day and weather effects (wind, solar, temperature), just as other conditional quantiles do. We therefore propose a model that allows the parameters of the tail distribution to depend on features that capture these effects.

We adopt a \gls{gamlss} \gls{gpd} model where the scale parameter is modelled via the link function $g_1(\cdot)=\log(\cdot)$,
with linear effects for 100m wind speed and surface solar irradiation, and a smooth effect for the estimated expected net-load, built using four basis functions.
The shape parameter $\xi$ is constant. 

\section{Evaluation Framework}
\label{sec:eval_framework}

This section introduces the methods and metrics we employ to evaluate and compare forecast performance. The objective of density forecasts it to be as sharp as possible while remaining reliable/calibrated~\cite{Gneiting2007a}. A sharp forecast is synonymous with low uncertainty and is therefore of more value to decision makers.
In addition to evaluating forecast based on this principle of `sharpness subject to calibration', we consider the impact on a key use-case for net-load forecasting in Section~\ref{sec:reserve}, namely reserve setting. 

\subsection{Pinball and Sharpness}


The Pinball Score for an individual quantile matches the loss function minimised in quantile regression model estimation. The Pinball Score is given by
\begin{equation}
    \frac{1}{T|\mathcal{A}|} \sum_{\alpha \in \mathcal{A}} \sum_{t=1}^T
 \left(q_{\alpha,t} - y_{t} \right)
 \left(\mathbb{1}(y_{t}\leq_{\alpha,t})-\alpha \right)
\end{equation}
where $\mathcal{A}$ is the set of quantiles being estimated. A drawback of the Pinball Score is that it places greater weight on the performance of quantiles close to $\alpha=0.5$ and less on those in the tails of predictive distributions. While Pinball Scores for individual tail quantiles may be calculated, they suffer from high variance due to the sparsity of observations and are therefore not suitable for discriminating between different forecasting systems. Therefore, we also evaluate interval width, the average of the $(1-2\lambda)\%$ prediction interval $q_{1-\lambda,t}-q_{\lambda,t}$. In both cases, however, calibration must be verified separately.

\subsection{Calibration}


Calibration, also called `reliability', is the property that forecast probabilities match the observed frequency of realisations. For example, if a forecast is calibrated, then 10\% of observations should fall below the $\alpha=0.1$ predictive quantile, with some tolerance based on the finite number of available forecast-observation pairs. This property is necessary for forecast probabilities to be used in quantitative decision making. Calibration is typically evaluated visually using reliability diagrams, which plot the nominal coverage, $\alpha$, against observer frequency mean($\mathbb{1}(y_{t}\le q_{\alpha,t})$).

Here we also examine so-called worm plots, a variant of the more familiar Q-Q plot, because they accentuate tail behaviours, which are of particular interest in this study~\cite{Buuren2001}. We calculate consistency intervals considering the temporal correlation of net-load, as the usual assumption of independence between samples does not hold: if the data lies within the consistency interval then the forecast is consistent with being calibrated~\cite{Pinson2010a}.

\section{Case Study}
\label{sec:case_study}

The proposed and benchmark methodologies described above are tested on a large dataset from \gls{gb} which is attached to this article as supplementary material~\cite{Browell2021_supZenodo}.
\blue{We define net-load as the load on the transmission system measured at the interface between transmission and distribution networks. In \gls{gb} these interfaces are called Grid Supply Points, and are grouped into 14 regions called \glspl{gspg}.} 
%
%
Net-load is metered at half-hourly resolution and our dataset spans five years 2014--2018. Over this period the installed capacity of embedded solar in GB grew from 3GW to 13GW, presenting a particular challenge.

Historic weather forecasts from the operational ECMWF-HRES model are used. We use forecasts with a base time of 0000h UTC which become available at approximately 0600h UTC and are therefore suitable for day-ahead forecasting tasks where a new forecast is issued in the morning of day $D$ for all of $D+1$. Forecasts of wind speed at 10m and 100m, surface temperature, cloud cover, solar irradiation and total precipitation are extracted on a $0.1^\circ$ spatial grid and pre-processed as described in Section~\ref{sec:weather_features}. The spatial extent of each \gls{gspg} is provided by National Grid ESO~\cite{NGESO_data2020} and population density by the Office for National Statistics~\cite{ONSGeography_data_2017,UK_population_2019}. We also include estimates of national embedded wind capacity from~\cite{NGESO_data2020}, estimates of regional solar capacity from~\cite{SolarSheffield_2019}, and public and school holidays.

 Regions vary in size from maximum net-load of 6.3GW in Group A (East England) to 0.5GW in Group P  (North Scotland), and minimum net-load of 1.9GW in Group C (Greater London) to significant export of -1.2GW in Group P. Prior to modelling and forecasting, all net-load data are standardised (converted to z-scores). A representative selection of results are presented here are in terms of standardised data for comparison between regions, in particular we highlight Groups C (city region), H (high solar) and P (high wind). Summary statistics of the dataset, the data itself and full results for all models and \glspl{gspg} are available in~\cite{Browell2021_supZenodo}.
 

Model development was carried out using three-fold cross-validation on data from 2014--2017 and final testing was performed on data from 2018. During the test phase \glspl{gam} were re-trained every two weeks in an expanding window fashion.

\subsection{Non-extreme Quantities}

\begin{figure}
     \centering
    \includegraphics[width=\columnwidth]{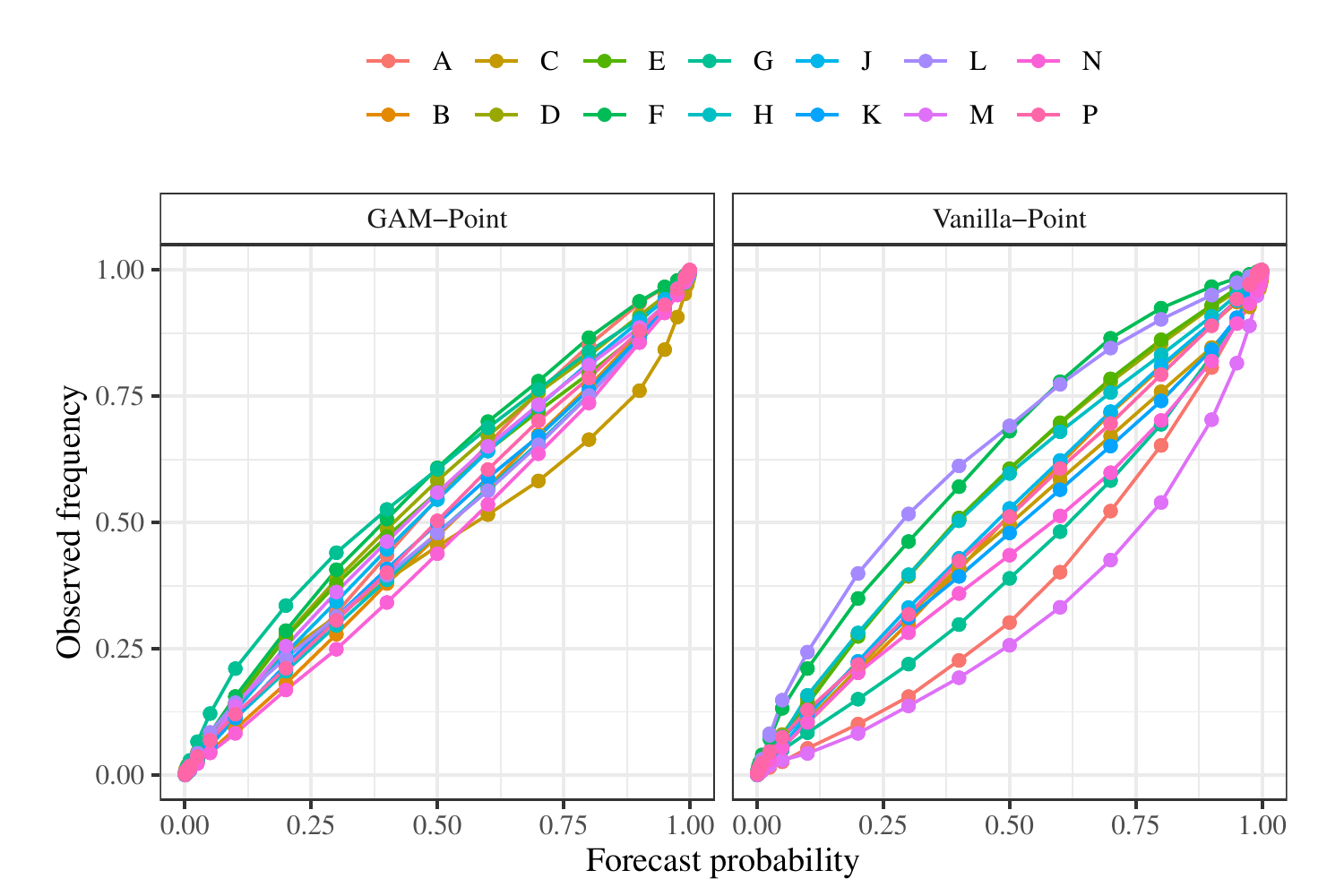}
        \caption{Reliability diagrams for GAM-Point and Vanilla-Point models and all \glspl{gspg} on the Test set.  If forecasts are calibrated the observed frequency of realisations falling below a given quantile $q_\alpha$ should match that quantile's forecast probability~$\alpha$.}
        \label{fig:Reliability}
\end{figure}

The central quantiles,  $\alpha=0.1,0.2,...,0.9$, are evaluated to verify their calibration and compare performance in terms of the Pinball Score. The evaluation of reliability of the Vanilla-Point and GAM-Point models for all 14 \glspl{gspg} is show in Figure~\ref{fig:Reliability}. Both models have near perfect calibration of the cross-validation exercises (shown in~\cite{Browell2021_supZenodo}), which degrades a little for GAM-Point on the final test dataset but considerably more for Vanilla-Point. In the case of the latter's forecasts, approximately half of the \glspl{gspg} are clearly not calibrated, highlighting the value of the more flexible GAM modelling approach. This is the pattern for all GAM- and Vanilla- models (shown in~\cite{Browell2021_supZenodo}). 


\begin{figure}
    \centering
    \includegraphics[width=\columnwidth]{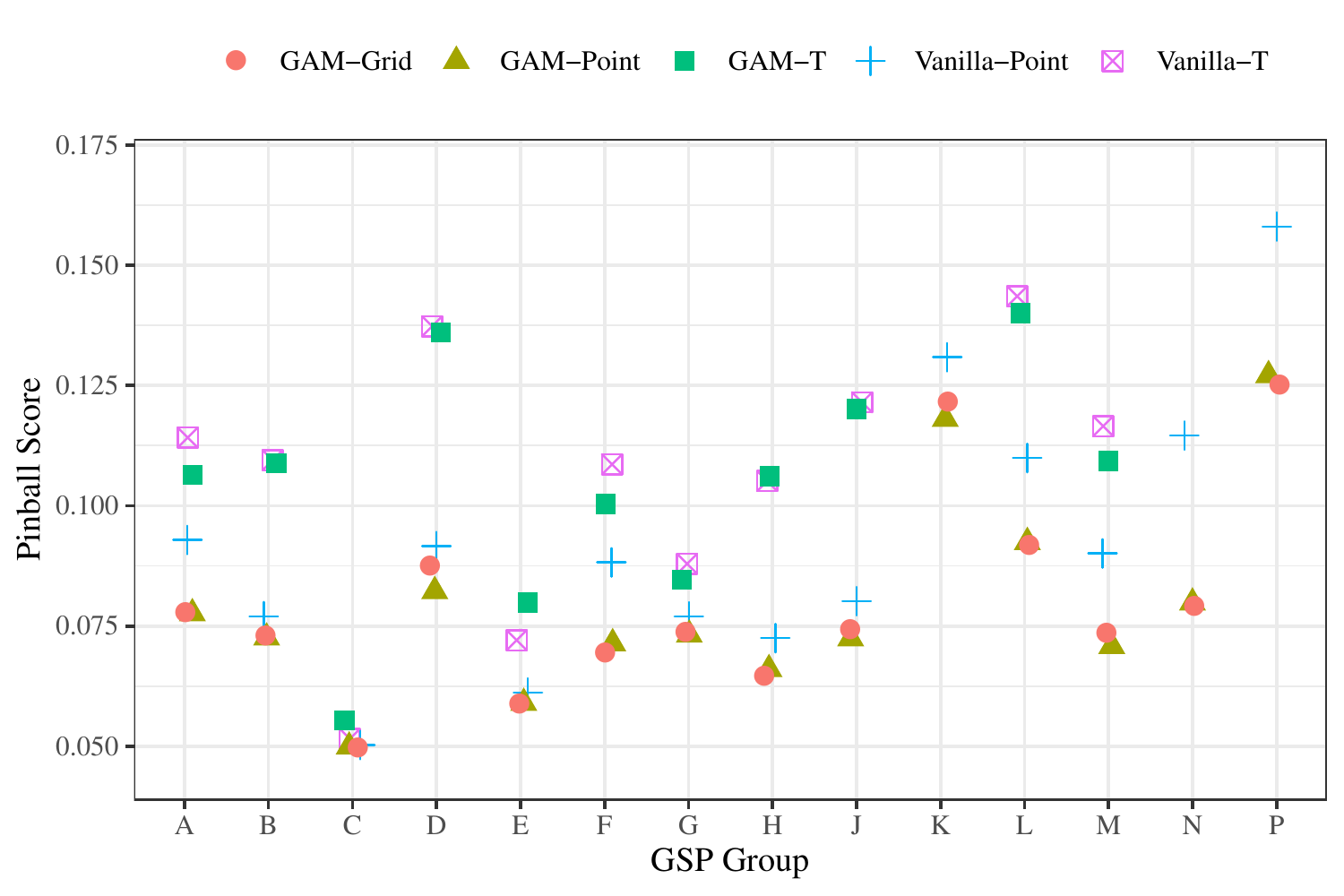}
    \caption{Pinball Score for each of the 14 \glspl{gspg} form the cross-validation exercises averaged across quantiles $\mathcal{A}=\{0.1,0.2,...,0.9\}$. The score is calculated on standardised data so is unitless and comparable between regions of different size. \blue{Scores for GAM-T and Vanilla-T exceed 0.2 at \glspl{gspg} K, N and P are not shown.}}
    \label{fig:Pinball}
\end{figure}

The Pinball Score for all models and \glspl{gspg} reveals similarly that the GAM approach results in superior forecasts compared to the Vanilla benchmarks, shown in Figure~\ref{fig:Pinball} and Table~\ref{tab:Pinball}. Furthermore, the importance of weather features related to embedded wind and solar generation are clear: Vanilla-T and GAM-T, the models with date/time and temperature features only, have extremely poor performance relative to their equivalents with additional wind and irradiance features. Comparing GAM-T and GAM-Point, the inclusion of additional weather forecast data results in a reduction of the Pinball Score by 40\% overall, and the effect is more prominent the greater the capacity of embedded generation. For example the improvement is only 10\% for Group C, the Greater London city region where there is little embedded generation but 60\% for Group P, where the embedded wind capacity exceeds peak load. A small improvement from addition of features derived from gridded NWP is suggested by the cross-validation exercise but is not consistently reproduced on test data and therefore inconclusive (Not shown, see supplementary material for details~\cite{Browell2021_supZenodo}).

\begin{table}
\caption{Pinball scores for cross-validation (CV) and Test exercises averages across all \glspl{gspg} (All), and for Groups C, H and P. The score for the best performing model(s) in each case are emboldened. \blue{Where the difference between scores of the two best methods is not significant at the $p<0.05$ level both are emboldened or italicised.}}
\label{tab:Pinball}
\centering
\begin{tabular}{l|c|c|ccc}
\hline
   & CV & \multicolumn{4}{c}{Test} \\ \hline
 GSP Group & All & All & C & H & P \\ 
  \hline
Vanilla-T & \textit{0.121} & \textit{0.137} & \textit{0.052} & \textit{0.105} & \textit{0.334} \\ 
  Vanilla-Point & 0.083 & 0.092 & \textbf{0.050} & 0.072 & 0.158 \\ 
  GAM-T & \textit{0.120} & \textit{0.134} & \textit{0.055} & \textit{0.106} & \textit{0.324} \\ 
  GAM-Point & \textbf{0.070} & \textbf{0.079} & \textbf{0.050} & 0.066 & \textbf{0.127} \\ 
  GAM-Grid & \textbf{0.070} & \textbf{0.080} & \textbf{0.050} & \textbf{0.065} & \textbf{0.125} \\ 
   \hline
\end{tabular}
\end{table}

\begin{figure}
    \centering
    \includegraphics[width=\columnwidth]{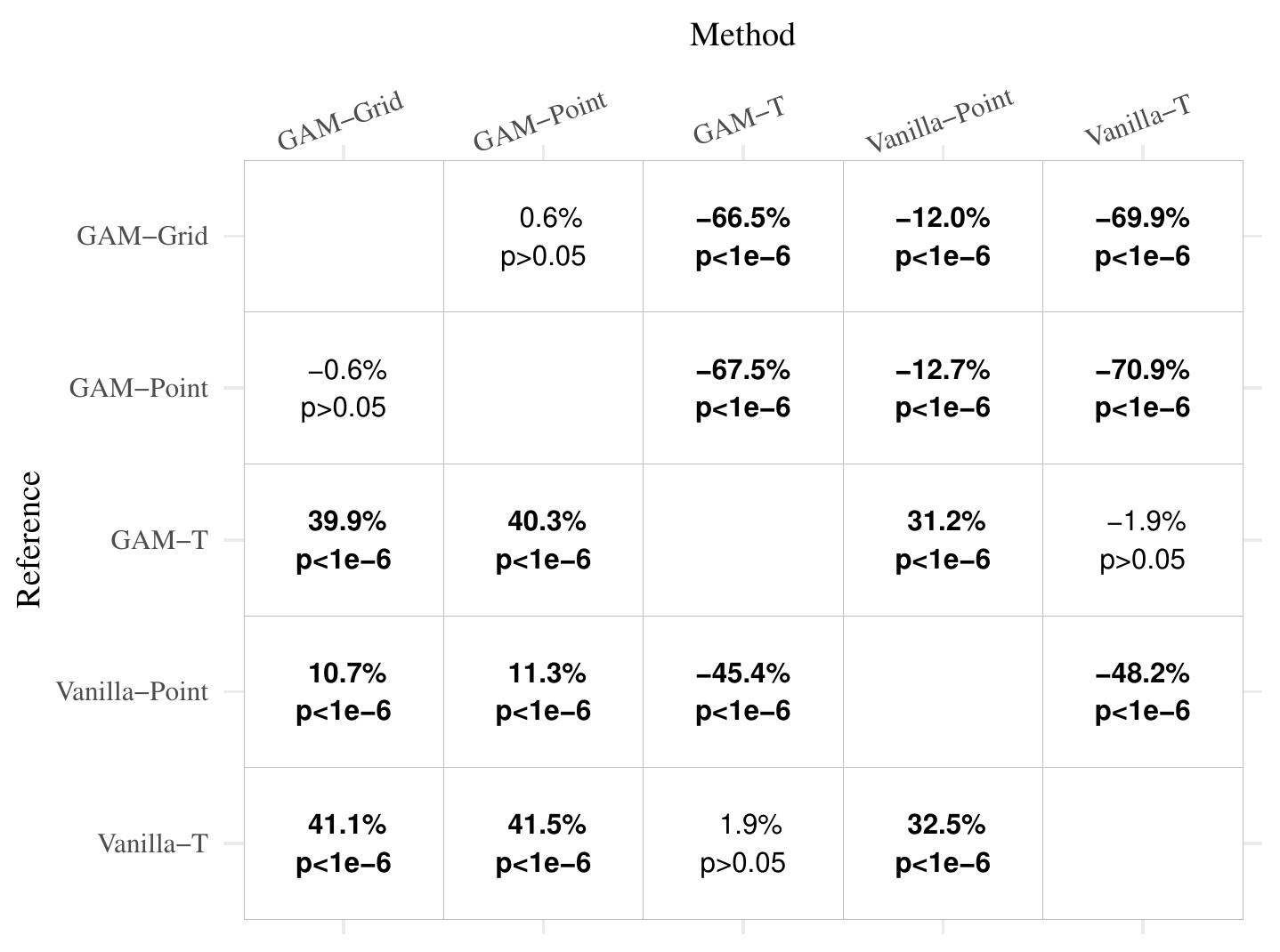}
    \caption{\blue{Pinball skill scores and significance ($p$-values from the Diebold-Mariano test~\cite{Diebold1995}) for the all pairs of methods, computed on the test set. A value of $p<0.05$ indicates strong evidence that the skill of forecast produced by `Method' relative to `Reference' is not zero. These value are emboldened.}}
    \label{fig:DMtest}
\end{figure}

\blue{The significance of apparent differences in skill is examined using the Diebold-Mariano (DM) test~\cite{Diebold1995}. For each \gls{gspg} and pair of models we have calculated the Pinball Skill Score and associated $p$-value from the DM test. The result for average performance across all \glspl{gspg} is shown in Figure~\ref{fig:DMtest}. The $>10\%$ improvement of GAM-Point/-Grid over the Vanilla-Point benchmark is significant ($p<0.001$), and this is consistent across all \glspl{gspg}. However, the skill of GAM-Grid relative to GAM-Point is more complex. Skill is positive and significant for \glspl{gspg} F and H, 
significant and negative for Groups D, J and K, and
not significantly different at \glspl{gspg} the remaining nine (shown in \cite{Browell2021_supZenodo}).}

\subsection{Extreme Quantities}

Extreme quantiles have been estimated using quantile regression in the same fashion as non-extreme quantiles for probability levels 0.05, 0.025, 0.01, 0.005, 0.0025, 0.001 and 0.0005, (and equivalent for right tail).
Evaluation of their reliability in the cross-validation exercise identified the $\alpha_L=0.025$ and $\alpha_R=0.975$ (0.05 and 0.95, respectively, for \glspl{gspg} E, F, L and P) as the last reliable quantiles in the present case study, so these are the probability levels beyond which \gls{gpd} tails are used as an alternative approach to estimating extreme quantiles. We consider both a static \gls{gpd} with constant parameters, and a conditional \gls{gpd} (\gls{gamlss}-type) where the scale parameter is a function of covariates and the shape parameter is constant, described in Section~\ref{sec:conditional_extremes}.

\begin{figure*}
     \centering
\begin{subfigure}[b]{0.32\textwidth}
         \includegraphics[width=\columnwidth]{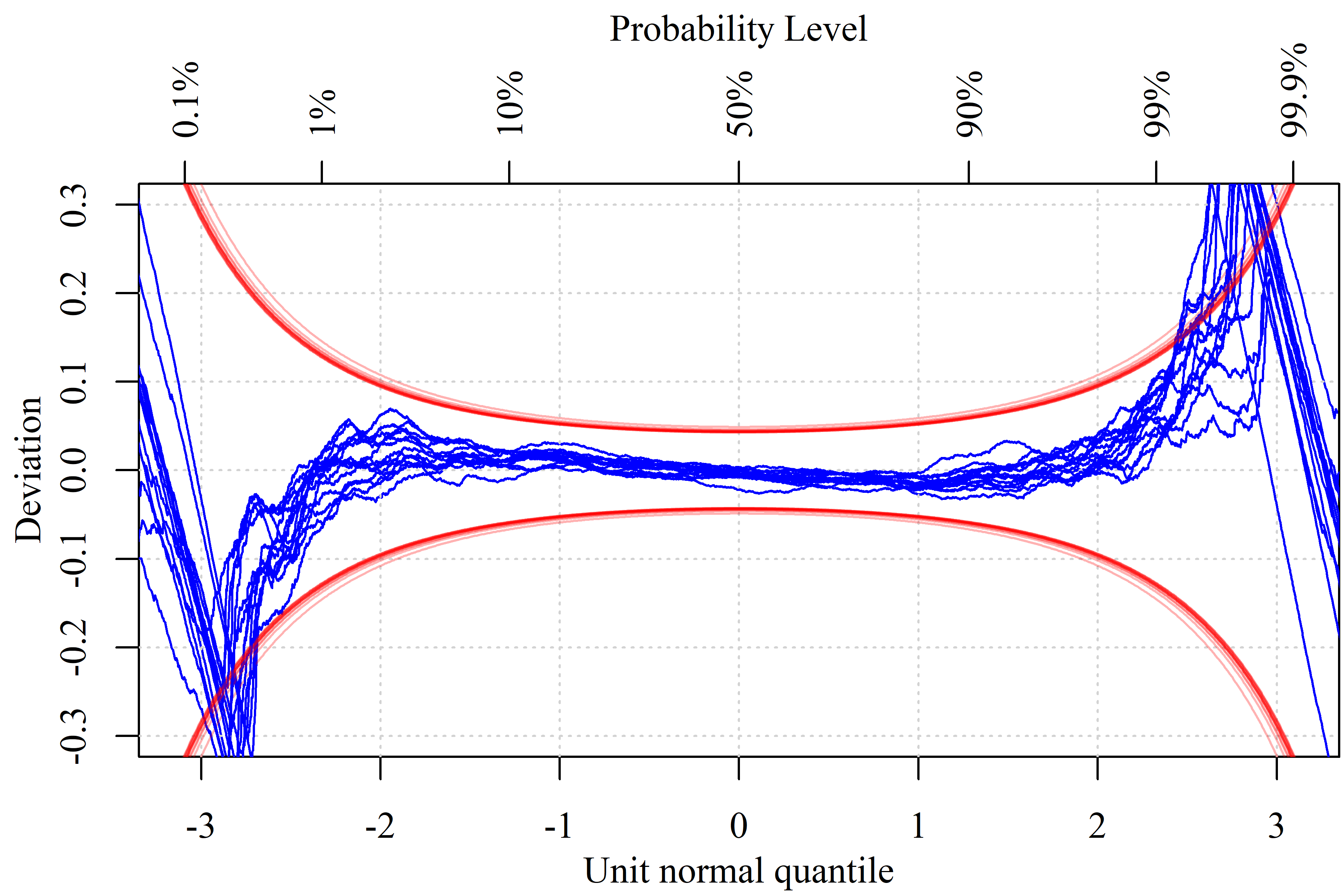}
    \caption{Quantile Regression}
    \label{fig:Worm_QR}
     \end{subfigure}
     \begin{subfigure}[b]{0.32\textwidth}
         \includegraphics[width=\columnwidth]{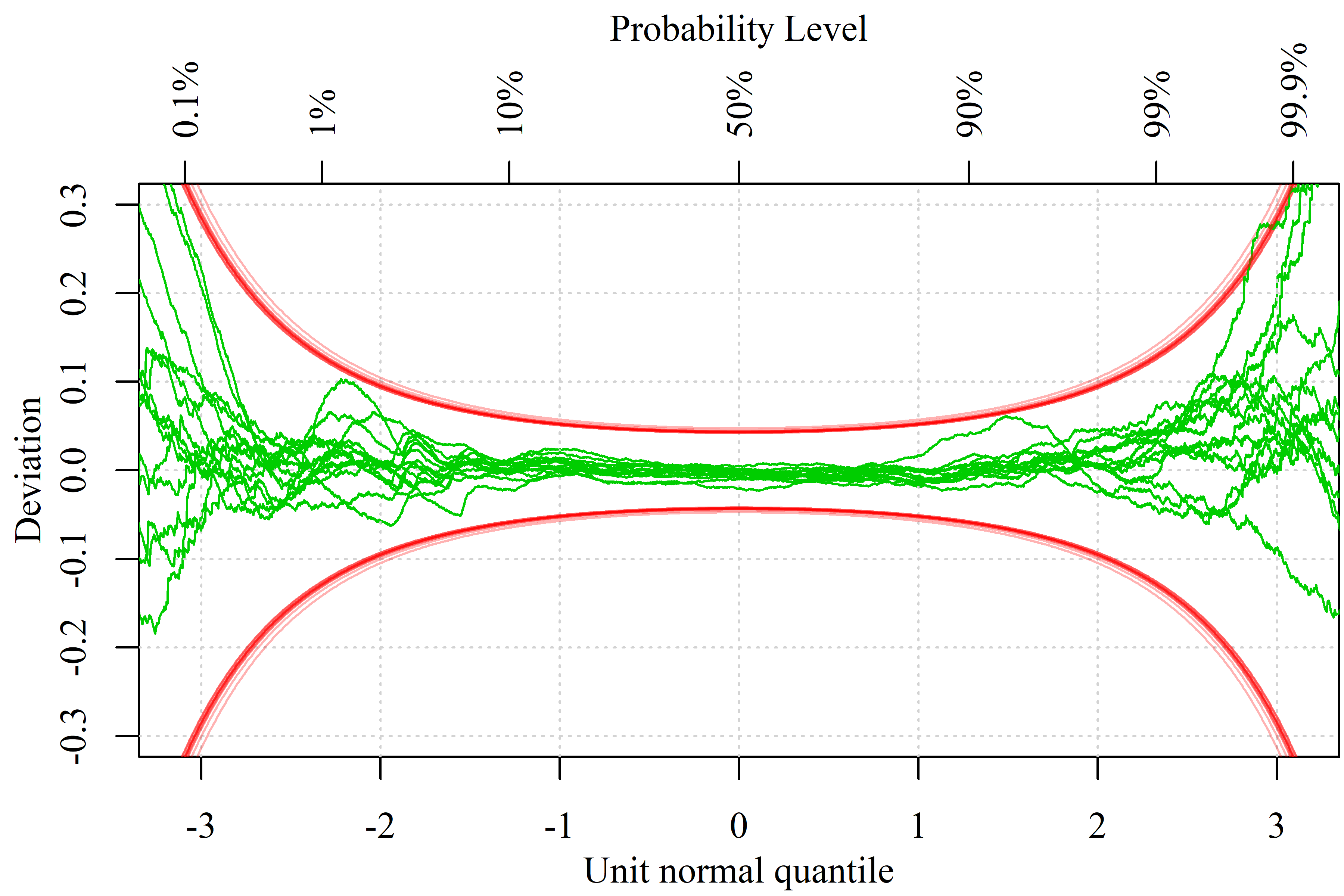}
    \caption{Static GPD}
    \label{fig:Worm_sGPD}
     \end{subfigure}    
     \begin{subfigure}[b]{0.32\textwidth}
         \centering
         \includegraphics[width=\columnwidth]{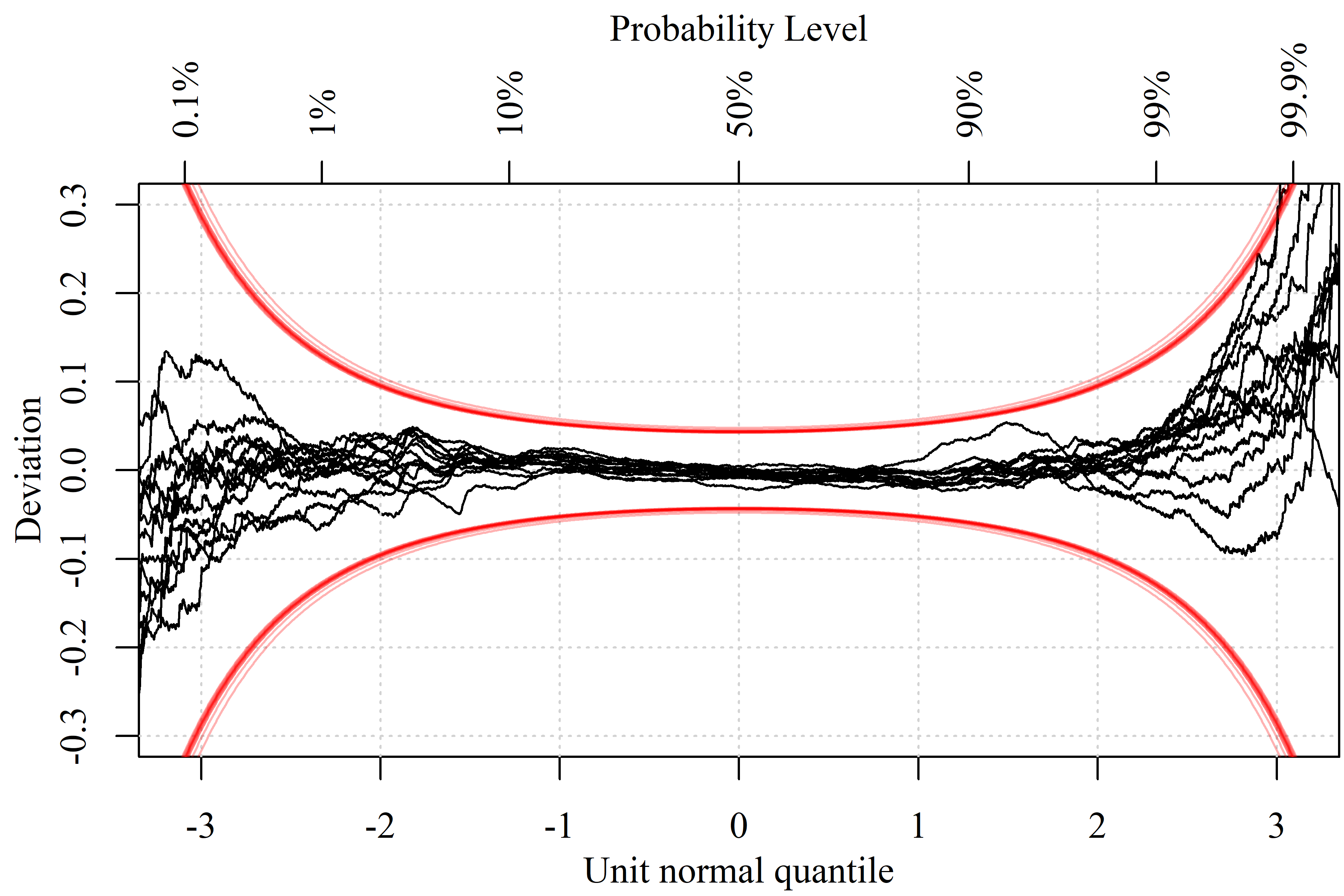}
    \caption{Conditional GPD}
    \label{fig:Worm_cGPD}
     \end{subfigure}
        \caption{Worm plots for GAM-Grid forecasts for 14 GSP Groups with  quantile regression tails (left), static (centre) and conditional GPD tails (right). Consistency intervals are calculated for each GSP Group at the 95\% level considering first-order auto-correlation and are overlaid in red. Forecasts that breach the consistency intervals are \textit{not consistent with being calibrated} and therefore not suitable for use.}
        \label{fig:WormPlots}
\end{figure*}

The worm plots in Figure~\ref{fig:WormPlots} evaluate the calibration of the predictive distributions, specifically the forecasts may be said to be `consistent with calibration' if the lines are within the consistency intervals. This reveals that below 1\% and above 99\% the forecasts based on quantile regression only are not calibrated at any \gls{gspg}. Therefore, these quantiles are not suitable for use in decision-making. However, those based on a combination of quantile regression and \gls{gpd} tails are consistent with calibration.

\begin{figure}
    \centering
    \includegraphics[width=\columnwidth]{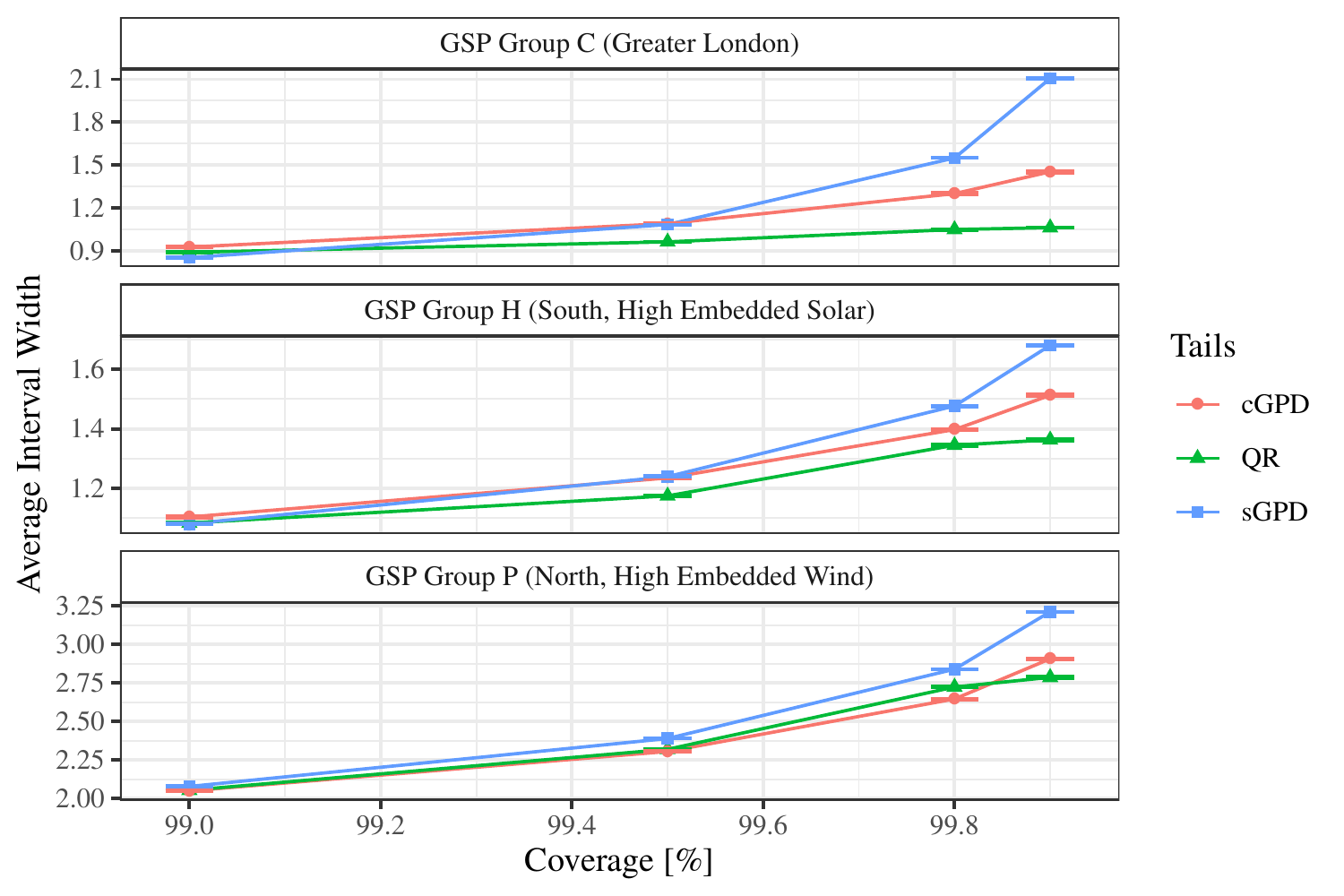}
    \caption{Sharpness diagrams for the GAM-Grid model with different approaches for estimating extreme quantiles. While quantile regression (QR) produces sharper intervals they are not calibrated so cannot be used. Of the two parametric approaches, which are calibrated, the conditional GPD (cGPD) produces sharper intervals than the static (sGPD). \blue{Bars are 95\% confidence intervals calculated via bootstrap re-sampling. They illustrate sampling variation and verify the significance of differences in interval width.}}
    \label{fig:Sharpness}
\end{figure}

We have seen that both static and conditional \gls{gpd} approaches produce calibrated forecasts, and are now interested in whether the relative complexity of the conditional models produce sharper forecasts. Average interval width is a simple measure of sharpness and visualised in Figure~\ref{fig:Sharpness} for GAM-Grid forecasts with quantile regression, static \gls{gpd} and conditional \gls{gpd} tails for \glspl{gspg} C, H and P. The 99.5\%--99.9\% interval forecasts using the conditional \gls{gpd} are much sharper than those from the static version, with more pronounced difference at higher probabilities. The conditional \gls{gpd} is therefore proving more confident forecasts than the static version, on average, while remaining calibrated. The practical value of this improvement is investigated in the next section. The intervals produced by quantile regression are sharper still, but as we have seen are not calibrated and therefore not usable.

\subsection{Use-case Evaluation: Reserve}
\label{sec:reserve}

Reserve is procured as a contingency to correct potential energy imbalances and the illustrative example here is based on current practice in \gls{gb}. The amount to procure is chosen to satisfy the TSO's risk appetite, e.g. reserve will be sufficient in all but 4 hours-per-year, which corresponds to the $\alpha= 0.0005$ quantile of net-load for upwards reserve (and $\alpha= 0.9995$ for downward). In this case we would calculate the upward reserve requirement in each 30-minute period to be $q_{0.5}-q_{0.0005}$ on the basis that the power system is scheduled according to the median forecast $q_{0.5}$.

In order to understand the impact of forecast performance on this task, we compare the volumes of reserve that would be procured using different forecasts. As well as total volumes, we are also concerned with the times of day when one forecast suggests procuring additional reserve because uncertainty is deemed high, compared to an alternative which is unable to discriminate between periods relatively high or low uncertainty.
Furthermore, as an additional benchmark we consider the naive approach of procuring reserve volumes using the empirical quantiles of historic deterministic forecast errors rather than a specific probabilistic forecast. This approach implicitly assumes that forecast uncertainty is constant, but it is easy to interpret and similar to the heuristics based on \textit{ex-post} analysis widely used in practice.

The practical benefits of the  performance gain offered by forecasts with the conditional over static \gls{gpd} tails discussed above translates into reduced volumes of both upwards and downward reserve. The former is reduced by up to 24.6\% and the latter by up to 10.8\% for probability levels 0.01\%--0.25\%, with greater impact at more extreme quantiles, see Table~\ref{tab:reserve}. The improvement is much greater when compared to the naive approach, especially at less extreme probability levels where the naive approach is particularly conservative.

\begin{table}
\centering
\caption{Percentage change in reserve volumes, and proportion of Periods (30 minute settlement periods) with lower reserve volumes, for forecasts based on conditional GPD tails relative to the naive and static GPD forecasts.}
\label{tab:reserve}
\begin{tabular}{l|l|rr|rr}
  \hline
  & & \multicolumn{2}{c}{cGPD vs Naive $[\%]$} & \multicolumn{2}{|c}{cGPD vs sGPD $[\%]$} \\
Reserve & Quantile & Volume & Periods & Volume & Periods \\ 
  \hline
\multirow{4}{*}{Upward} &0.01\% & -18.0 & 77.0 & -24.6 & 81.3 \\ 
  & 0.05\% & -16.5 & 73.6 & -13.9 & 70.4 \\ 
  & 0.1\% & -17.2 & 74.1 & -9.1 & 63.3 \\ 
  & 0.25\% & -16.5 & 71.8 & -3.2 & 55.9 \\   \hline
  \multirow{4}{*}{Downward} &  99.75\% & -19.8 & 81.4 & -0.8 & 59.7 \\ 
  & 99.9\% & -22.1 & 84.0 & -3.2 & 64.1 \\ 
  & 99.95\% & -22.9 & 84.7 & -5.3 & 67.3 \\ 
  & 99.99\% & -25.8 & 85.6 & -10.8 & 74.3 \\
   \hline
\end{tabular}
\end{table}

\begin{figure}
    \centering
    \includegraphics[width=\columnwidth]{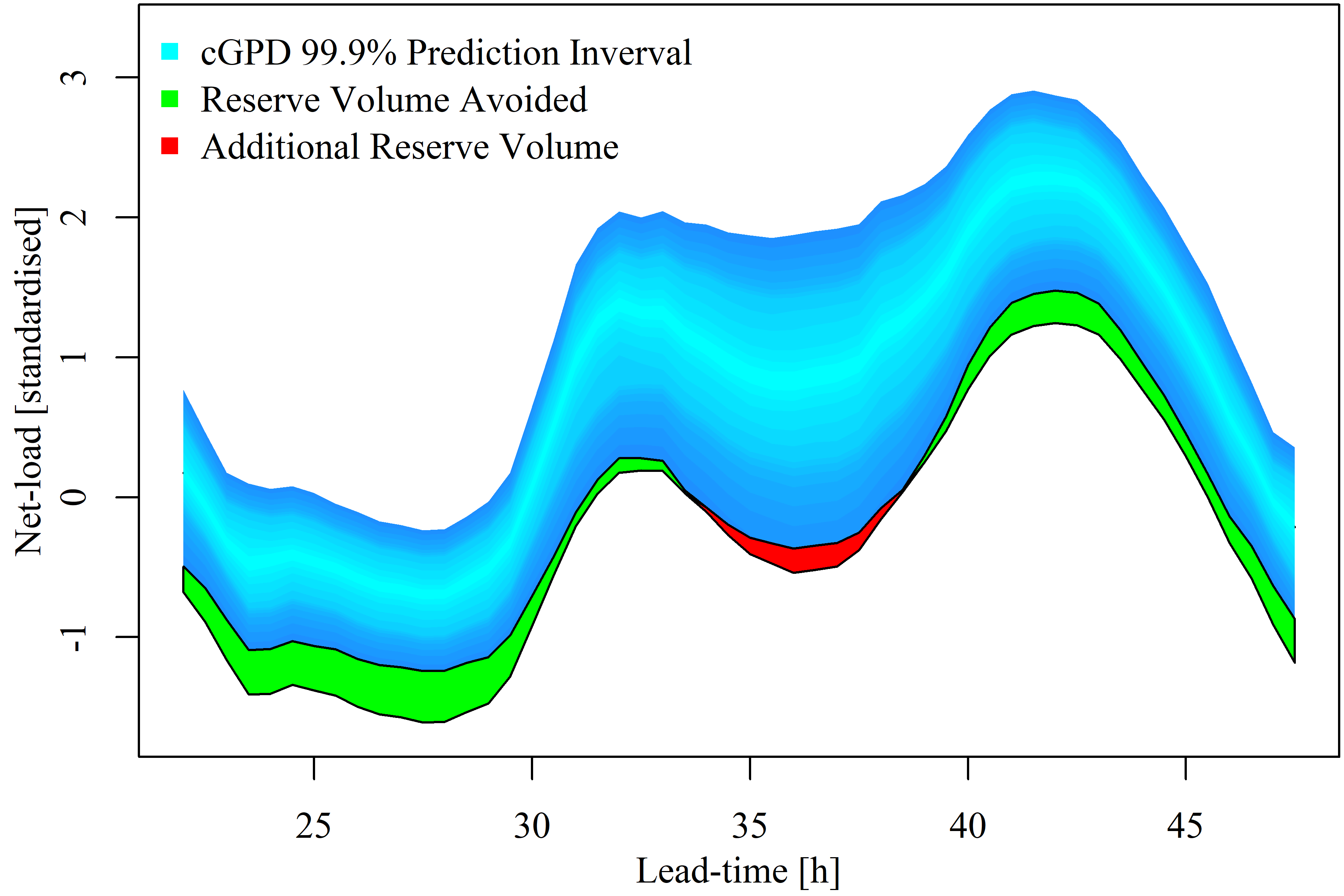}
    \caption{Illustration of the difference in reserve holding at 0.05\% level between static and conditional GPD at \gls{gspg} H. Compared to the sGPD, the cGPD requires less reserve to be held most of the time to maintaining the same risk level (green), but more reserve during period of greater uncertainty which the sGPD does not capture (red). Lead-time is hours since 2018-12-09T00:00:00Z.}
    \label{fig:reserve}
\end{figure}

Critically, the reduction in reserve is not uniform over time. Procurement based on the cGPD requires less upward reserve than the sGPD in 56\%--81\% of settlement periods (depending on probability level), meaning it requires more reserve in 19\%--44\% of periods to meet the same risk level. The cGPD, being conditioned on weather effects, captures variation in uncertainty based on these features. In some time periods the forecast is relatively sharp, (more confident, less reserve required), in others it is less sharp (greater uncertainty, more reserve required). This effect is visualised in Figure~\ref{fig:reserve}, where the conditional \gls{gpd} captures reduced uncertainty in the early hours and greater uncertainty during the middle of the day and solar peak, relative to the static \gls{gpd}, which, loosely speaking, `averages' these effects over time. This is important, as without this information the \gls{tso} may hold more reserve than necessary some of the time, incurring a cost, and not enough to satisfy their risk appetite at other times, exposing themselves to excess risk. This effect is more pronounced for the naive approach, which does not capture any variation in uncertainty at all (the sGPD forecasts captures some via the last conditional quantiles $q_{\alpha_L}$ and $q_{\alpha_R}$). 

\section{Discussion and Conclusions}
\label{sec:discussion}

Predicting the possible extremes of future net electricity load is of interest to power system operators who must manage associated risks. However, this is challenging due to the inherent lack of data when modelling extremes, and the complexity introduced by embedded renewables.
In this paper we have shown that the Generalised Pareto Distribution with a time-varying scale parameter can accurately describe the tails of predictive distributions. Time-variation is a result of conditioning the scale parameter on expected load and weather forecasts. The resulting forecasts are calibrated, unlike those produced by quantile regression, and sharper than those obtained using a non-time-varying approach.

Extensive evaluation has been performed based on data from the 14 regions of Great Britain's electricity system, which exhibit a diversity of embedded wind and solar capacities relative to load. In addition to statistical measures, the practical task of reserve energy procurement to meet a given risk level has been considered. Conditioning the \gls{gpd} allows forecasts to reflect changing uncertainty with the overall effect of reducing the volume of reserve required to achieve the target risk level. Importantly, while the overall volume of reserve is reduced relative to benchmarks by using the conditional \gls{gpd}, during some periods reserve volume is higher, implying that the benchmarks expose users to excessive risk during these periods. The proposed method has the dual benefit of reducing cost and exposure to risk.

As part of this work we also studied the potential benefits of deriving weather features from gridded \gls{nwp} to improve short-term net-load forecasts, as has been demonstrated for individual wind and solar farms. Gridded \gls{nwp} does not appear to add significant value to deterministic and probabilistic net-load forecasts in the present framework, but it is possible that other forecasting methods would be able to extract value from this data \blue{by constructing different features, for example}.

\blue{We have developed a complex model for net-load forecasting in \gls{gb} with some \gls{gb}-specific date/time features, however, the value of wind and solar-related weather features for net-load forecasting will be universal wherever the capacity of embedded wind and/or solar is significant. The model structure developed here, GAM with quantile regression and conditional parametric tails, should also generalise well to other applications. The problem of describing the tails of probabilistic forecasts that depend on covariates is certainly not unique to net-load forecasting, similar problems exist in meteorology, finance and logistics, for example.}

Other future work should consider the spatio-temporal dependency between regions, and important aspect of energy system operation~\cite{Martinez2018}, and how this may depend on weather effects as the tails of predictive distribution do. Alternative approaches to predicting extremes of probabilistic net-load forecasts based on decomposing net-load into its constituent parts could also be considered. but would require careful modelling of potential tail dependency structures.

\section*{Acknowledgements}

Jethro Browell is supported by EPSRC Fellowship (EP/R023484/1) and a visiting position at the University of Bristol as Heilbronn Visitor in Data Science in February 2020.
The authors thank National Grid ESO for many discussions on forecasting and reserve setting,  Graeme Hawker for support accessing GSP data, Ciaran Gilbert for contributions to \textit{ProbCast}~\cite{Browell2020_probcast}, and the anonymous editor and reviewers who helped us improve this paper.
\textbf{Data statement:} The supplementary material attached to this paper includes data generated by this research and code to reproduce it~\cite{Browell2021_supZenodo}.

\ifCLASSOPTIONcaptionsoff
  \newpage
\fi



\bibliographystyle{IEEEtran}
\bibliography{MyEndNoteLibrary}
%



%

\begin{IEEEbiography}[{\includegraphics[width=1in,height=1.25in,clip,keepaspectratio]{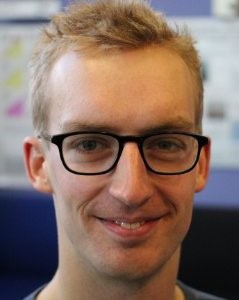}}]{Jethro Browell}
received the M.Phys. degree in mathematics and theoretical physics from the University of St Andrews, U.K., in 2011, and the Ph.D. degree in wind energy systems from The University of Strathclyde, U.K., in 2015. He is a Senior Lecturer in the University of Strathclyde’s Institute for Energy and Environment where his research interests span all aspects of energy forecasting and associated decision-making in power system operation and energy markets.

Dr. Browell is an EPSRC Innovation Fellow, a member of the IEEE Power \& Energy Society and Vice-chair of the PES Working Group on Energy Forecasting and Analytics, and an editorial board member for \textit{Renewable and Sustainable Energy Reviews} and \textit{Sustainable Energy, Grids and Networks}.
\end{IEEEbiography}

\begin{IEEEbiography}[{\includegraphics[width=1in,height=1.25in,clip,keepaspectratio]{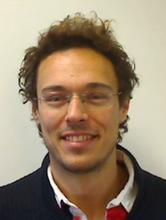}}]{Matteo Fasiolo} 
received an M.Eng. degree in Industrial Engineering from the University of Udine, Italy, in 2010, an M.Sc in Financial Engineering from Birkbeck, University of London, U.K., in 2011, and a Ph.D. degree in Statistics from the University of Bath, U.K., in 2016. He is a Lecturer in the University of Bristol’s Institute for Statistical Science, where he works on developing non-parametric regression modelling methodology, with a particular focus on energy-related applications.
\end{IEEEbiography}







\end{document}